# Fragile "Symmetry-Protected" Half Metallicity in Two-Dimensional van der Waals Magnets: A Case Study of Monolayer FeCl$_2$

Qiushi Yao[1], Jiayu Li[1] and Qihang Liu[1,2,3,*]

[1]*Shenzhen Institute for Quantum Science and Engineering and Department of Physics, Southern University of Science and Technology, Shenzhen 518055, China*

[2]*Guangdong Provincial Key Laboratory of Computational Science and Material Design, Southern University of Science and Technology, Shenzhen 518055, China*

[3]*Shenzhen Key Laboratory of for Advanced Quantum Functional Materials and Devices, Southern University of Science and Technology, Shenzhen 518055, China*

[*]Email: liuqh@sustech.edu.cn

Abstract

Two-dimensional (2D) half-metallic materials are of great interest for their promising applications in spintronics. Although numerous of 2D half-metals have been proposed theoretically, rarely of them can be synthesized experimentally. Here, exemplified by monolayer FeCl$_2$, we show three mechanisms in such quantum magnets that would cause the metal-insulator transition by using first-principles calculations. In particular, half-metallicity, especially that protected by symmetry-induced degeneracies, predicted by the previous theoretical simulations could be destroyed by electron correlation, spin-orbit coupling and further structural distortions to lower the total energy. Our work reveals the fragility of the symmetry-protected half-metals upon various competing energy-lowering mechanisms, which should be taken into account for theoretically predicting and designing quantum mateirals with exotic functionalities.

# I. INTRODUCTION

Two-dimensional van der Waals (2D vdW) magnetic materials exfoliated from their bulk counterparts have gained extensive attention since the successful synthesis of atomically thin $CrI_3$ [1], $Cr_2Ge_2Te_6$ [2] and $Fe_3GeTe_2$ [3] flakes. These 2D vdW magnets with intrinsic magnetic properties provide ideal platform for exploring the long-range magnetic order in the 2D limit, as well as show promising applications in next-generation spintronic devices. Practical applications of 2D vdW magnets require considerable magnetocrystalline anisotropy energy (MAE) and room-temperature $T_C$, which are the main focuses of most previous studies [3,4]. Yet, another typical character of these 2D vdW magnets is that most of them are $3d$ transition metal compounds, which provides unique opportunities for studying the coupling between magnetism (Hund's interaction as well as Jahn-Teller effect) and electron correlation. Such phenomena have been widely studied in perovskites, leading to a plethora of exotic effects featured by charge, spin and orbital degrees of freedom [5-7]. However, only rarely are such phenomena mentioned in the present literature devoted to 2D vdW magnets [8-10]. Furthermore, from $3d$ to $4d$ and $5d$ transition metal 2D vdW magnets, albeit larger crystal field splitting and weaker Hund's interaction entail them preferring low-spin states, which suppressed the magnetism, strong intrinsic spin-orbit coupling (SOC) brings about interesting and nontrivial effects and acts as another knob to tune the properties of 2D vdW magnets, resembling the well-known iridate $Sr_2IrO_4$ [11]. In addition to the providing of MAE that harbors 2D magnetism, how SOC will affect the properties of 2D vdW magnets is another interesting and fundamental problem.

Very recently, a new 2D vdW magnet, i.e. $FeCl_2$, has been synthesized by molecular beam epitaxy [12,13]. With a nominal valent state of +2, each octahedrally coordinated $Fe^{2+}$ ion holds six $d$-electrons with the population of $t_{2g}^{3\uparrow}e_g^{2\uparrow}$ and $t_{2g}^{1\downarrow}$, possessing high-spin state [12,13]. Theoretical studies predicted monolayer $FeCl_2$ as a SOC-assisted Mott insulator with the spin-down electron occupying a nondegenerate orbital guaranteed by local trigonal distortion (TD) and SOC [14]. However, many other theoretical studies predicted a half-metallic state with large spin gap [12,15-19]. Indeed, *Cai* et al. demonstrated that monolayer $FeCl_2$ monolayer is a uniform insulator with a gap in the magnitude of an electronvolt [13]. However, the authors didn't exclude the possible half-metallicity of the freestanding monolayer $FeCl_2$ with the arguments that the insulative nature may be induced by the extra electrons from the substrate, leading the Fermi level moves up and enters the band gap [13]. To the best of our knowledge, unintentionally controlled elections injection decorating "half-metallic" $FeCl_2$ to be insulating is somewhat not that ready. Controversies between various studies call for prompt theoretical and experimental verifications.

In this work, exemplified by monolayer $FeCl_2$, we endeavor to obtain a microscopic insight into the discrepancy between theoretical and experimental studies. By employing density functional theory (DFT) calculations, we show that the predicted "robust" half-metallicity of monolayer $FeCl_2$ is actually guaranteed by the coexistence of orbital ordering (lower-lying $e'_g$ and higher-lying $a_{1g}$) and symmetry-protected double degeneracy of $e'_g$. However, such half-metallicity tends to be destroyed by considering some realistic symmetry-breaking mechanism, *i.e.*, SOC and structural distortions. Furthermore, we find that electron correlation can switch the $a_{1g}$-$e'_g$ relative order, leading to metal-insulator transition of monolayer $FeCl_2$. Our work not only reveals the fragility of the exotic symmetry-protected half-metals upon energy-lowering mechanisms, but also sheds light on the rational ground state prediction by avoiding local minima in first-principles calculations.

**II. COMPUTATIONAL METHODOLOGY**

First-principles calculations were carried out using Vienna *ab initio* simulation package (VASP) [20] within the framework of density functional theory (DFT) [21,22]. Exchange-correlation functional was described by generalized gradient approximation with the Perdew-Burke-Ernzerhof (PBE) formalism [23]. The electron-ion interaction was treat by projector-augmented-wave (PAW) potentials [24] with a planewave-basis cutoff of 500 eV. The whole Brillouin-zone was sampled by $15 \times 15 \times 1$ Monkhorst-Pack grid [25] for monolayer $FeCl_2$. Due to the correlation effects of $3d$ electrons in Fe atoms, we employed GGA + $U$ approach within the Dudarev scheme [26]. A large scale of $U$ (0.1 – 4 eV) was applied to study the evolution of electronic properties with $U$. We took the experimental lattice constant, namely, a = b = 3.603 Å [27], and fixed it while structural optimization. (For comparison, we also optimized the lattice constant, which converged to 3.592 Å, in good consistent with the experimental value.) A vacuum space larger than 15 Å was employed to avoid artificial periodic image interactions between $FeCl_2$ layers. All atoms were fully relaxed until the force on each atom was less than 0.01 eV/Å and the total energy minimization was performed with a tolerance of $10^{-5}$ eV. A $4 \times 4 \times 1$ supercell was constructed and relaxed without any symmetry constrains to investigate further structural distortions. Brillouin-zone of supercell was sampled by $5 \times 5 \times 1$ Monkhorst-Pack grid. Open source code BandUP [28,29] was used to unfold the supercell band structure to get the effective band structure (EBS).

It was worth mentioning that in order to avoid converging to the local minima of the potential energy surface, we took two initial electron populations as the starting point, namely, $e_g'^1 a_{1g}^0 e_g^0$ and $e_g'^0 a_{1g}^1 e_g^0$, and then do self-consistent calculations including a full electronic relaxation (here we only considered possible occupation configurations of the single property-determining spin-down electron). Both populations were achieved by assuming initial electron occupation matrix via the open source software developed by Watson [30]. By this procedure, we can reliably get the ground state by a direct comparison of the two different configurations. Such procedure was necessary to determine the groundstate of monolayer $FeCl_2$ and has been widely used in previous studies [10,31-33]. For simplicity, we marked the state wherein the $d^{1\downarrow}$ electron occupies $e_g'$ doublet as $e_g'^1$ instead of $e_g'^1 a_{1g}^0 e_g^0$, and state wherein the $d^{1\downarrow}$ electron occupies $a_{1g}$ singlet as $a_{1g}^1$ to replace $e_g'^0 a_{1g}^1 e_g^0$. Open source software VESTA [34] and VASPKIT [35] were used to visualize and deal with VASP output files.

## III. RESULTS AND DISCUSSION

**Basic electronic structure.** As shown in Figure 1(a), monolayer $FeCl_2$ crystallizes in *T*-phase monolayer $MoS_2$ structure with the Fe-layer sandwiched between two Cl-layers. Each Fe atom is surrounded by six first-neighbor Cl atoms forming a $FeCl_6$ octahedron, which arranged in an edge sharing pattern. In the absence of any distortion, the octahedral crystal field splits Fe-3$d$ orbitals into the lower-lying triply degenerate $t_{2g}$ and the higher-lying doubly degenerate $e_g$ orbitals. Taking the 2D nature of the monolayer and the local compressed trigonal distortion into account (featured by $\theta > \theta_0 = 54.74°$) [10,14,36], the triply degenerated $t_{2g}$ orbitals are split into $a_{1g}$ singlet and $e'_g$ doublet, as schematized in Figure 1(b). As anticipated, the band structure calculated by the pure spin-restricted GGA calculations demonstrates the crystal field levels, as illustrated in Figure S1. The energy splitting between the entirely occupied $t_{2g}$ and empty $e_g$ manifolds is about 1.5 eV, while the local trigonal distortion induced gap between $a_{1g}$ and $e'_g$ levels is about an order of magnitude smaller, a hallmark of second-order effect compared with the octahedral crystal field splitting. Due to the similar energies between $a_{1g}$ and $e'_g$ manifolds, the $a_{1g}$-$e'_g$ relative order is more subtle, which depends on the crystal shape (compressed or elongated trigonal distortion), electron filling, $e'_g$-$e_g$ mixing, and long-ranged crystal field due to the lattice anisotropy [37,38]. The hierarchy of $a_{1g}$ and $e'_g$ levels is especially prominent for determining the ground state of monolayer $FeCl_2$, which will be detailed below (Mechanism I).

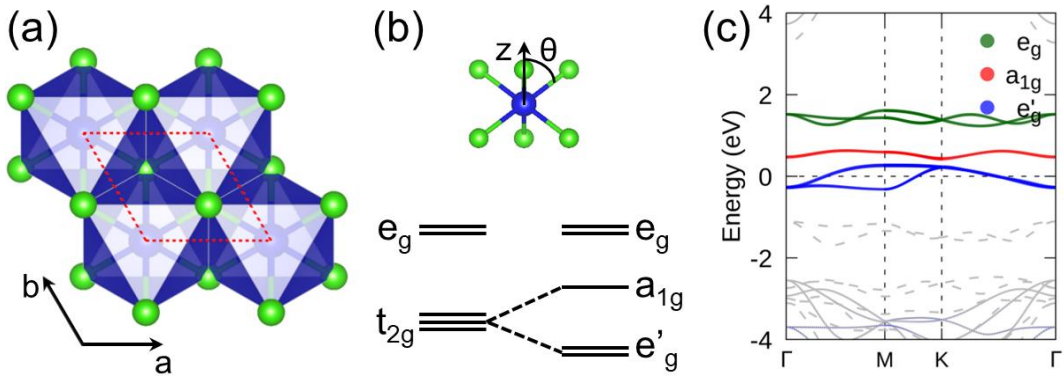

**Figure 1.** (a) Crystal structure of monolayer $FeCl_2$. Red dashed lines outline the unit cell. Blue and green spheres represent Fe and Cl atoms, respectively. The $FeCl_6$ octahedrons are colored by blue. (b) Compressed local trigonal distortion featured by $\theta > \theta_0$, where $\theta$ is the angle between the $z$-axis (here is the crystallographic $c$-axis) and Fe-Cl bonds. In undistorted octahedron, $\theta_0 = 54.74°$. Crystal field splitting diagram of $d$ orbitals induced by octahedral crystal field and trigonal distortion. For simplicity we show here $a_{1g}$ singlet is higher than $e_g'$ doublet, while the reversed order is also possible (see text for detail). (c) Band structure calculated by spin-unrestricted GGA method with initial electron population $e_g'^1 a_{1g}^0 e_g^0$, leading to the final $e_g'^1$ state. The spin-down Fe-3$d$ orbitals are highlighted by different colors. Spin-up and spin-down channels are indicated by dashed and solid lines, respectively.

Further considering the Hund's interaction, monolayer FeCl$_2$ possesses high-spin ground state with the electron population of $d^{5\uparrow} d^{1\downarrow}$, giving the total moment of 4 $\mu_B$/f.u. (see Figure 1(c)) [12,14-18]. Our calculations show that both initial starting points (see Method for details), $e_g'^1 a_{1g}^0 e_g^0$ and $e_g'^0 a_{1g}^1 e_g^0$, converge to final $e_g'^1$ configuration, as shown in Figure 1(c) and Figure S2. The spin-down $e_g'$ doublet encompasses the Fermi level, while the $a_{1g}$ singlet resides between $e_g'$ and $e_g$ states. In fact, at $\Gamma$ ($K$), the two $e_g'$ orbitals form the basis function $\{\frac{1}{\sqrt{3}}(d_{xy} + e^{i2\pi/3} d_{xz} + e^{i4\pi/3} d_{yz}), \frac{1}{\sqrt{3}}(d_{xy} + e^{i4\pi/3} d_{xz} + e^{i2\pi/3} d_{yz})\}$ (local coordinate) of a two-fold irreducible representation of the little group D$_{3d}$ (D$_3$), leading to symmetry-enforced double degeneracy. With only one electron occupies the doubly degenerated $e_g'$ orbitals, monolayer FeCl$_2$ exhibits intrinsic half-metallicity, consistent with the previous studies [12,15-19]. Such symmetry-protected half-metallicity is robust as long as the specific orbital ordering (lower-lying $e_g'$ and higher-lying $a_{1g}$) and symmetry (double degeneracy) coexist. However, as will be detailed below, the desired coexistence could be easily destroyed by electron correlation (Mechanism I), spin-orbit coupling (Mechanism II) and further symmetry-breaking energy-lowering distortions (Mechanism III). All these mechanisms tend to lift the degeneracy, resulting in metal-insulator transition of FeCl$_2$.

**Mechanism I: Electron correlation induced orbital switching.** Firstly, we carry out spin-unrestricted GGA+$U$ calculations to elucidate the electron correlation effect in a mean-field level. As we will see, moderate correlation can switch the $a_{1g}$-$e'_g$ ordering, leading to metal-insulator transition of FeCl$_2$. A wide range of $U$ (0.1 – 4 eV) are tested to ascertain the band evolution with $U$. On one hand, as can be seen from Figure 2(a) (also see Figure S3), starting from $e'^1_g$ occupation as the initial configuration for electron density, the band structures calculated with all $U$ results in half-metallic ground state. In fact, without any symmetry broken, any reasonable $U$ or other functional (such as meta-GGA and hybrid functional) cannot break the double degeneracy at Γ and $K$. The increase of $U$ only pushes $a_{1g}$ orbital far away from Fermi level but cannot split the $e'_g$ doublet at these high-symmetry momenta, leading to the seemingly "robust" half-metallic phase reported in the literature [12,15-19]. On the other hand, starting from $a^1_{1g}$, we get totally different results. Except for $U$ = 0.1 eV, from $U$ = 0.2 to 4 eV, all $e'^0_g a^1_{1g} e^0_g$ starting point can converge to $a^1_{1g}$, as shown in Figure 2(b) and S4. Even with $U$ as small as 0.2 eV (Figure S4(b)), the $a^1_{1g}$ electron population can induce a gap between $a_{1g}$ and $e'_g$ orbitals, leading to an insulator phase. Increasing $U$ from 0.2 to 4 eV (Figure 2(b) and S4), monolayer FeCl$_2$ maintains insulation with monotonically enhanced gap size because $e'_g$ orbitals are pushed away from the Fermi level by $U$. Unlike the partially occupied $e'_g$ orbitals in $e'^1_g$, bandwidth of the totally unoccupied $e'_g$ orbitals in $a^1_{1g}$ configuration is suppressed by $U$ evidently. At $U$ = 0.1 eV (Figure S4(a)), initial $a^1_{1g}$ distribution converges to $e'^1_g$ as the ground state (same as the case without $U$, see Figure S2). Potential surface at $U \leq 0.1$ eV is illustrated in the left panel of Figure 2(e), $a^1_{1g}$ turns out to be an instable configuration.

To determine the ground state between the two local minima for $U$ = 0.2 to 4 eV, we compare the total energies of these two states. As illustrated in Figure 2(c), with $U$ smaller than 0.8 eV, half-metallic $e'^1_g$ state is favored in energy, while insulating $a^1_{1g}$ state becomes more energetically favorable with larger $U$ (0.8 – 4 eV). For Fe, the reasonable $U$ would be in the range of 2 – 4 eV [39]. Thus, based on our calculations, it is reasonable to suggest that with moderate electron correlation, FeCl$_2$ favors the insulating ground state, which is confirmed by a recent experiment [13]. To obtain the insulating $a_{1g}^1$ state, moderate electron correlation and proper initial electron occupation matrix should be taken into account. As illustrated in the middle (right) panel of Figure 2(e), $a_{1g}^1$ ($e_g'^1$) configuration plays as a local minima of the potential surface while $e_g'^1$ ($a_{1g}^1$) configuration is ground state at $0.1 < U < 0.8$ eV ($U \geq 0.8$ eV). Another intriguing difference between $a_{1g}^1$ and $e_g'^1$ is that while $a_{1g}^1$ is a spin-only system ($l_z = 0$), $e_g'^1$ renders monolayer FeCl$_2$ a system with orbital degree of freedom, resulting in fruitful orbital physics under certain condition, especially after considering spin-orbit coupling (Mechanism II).

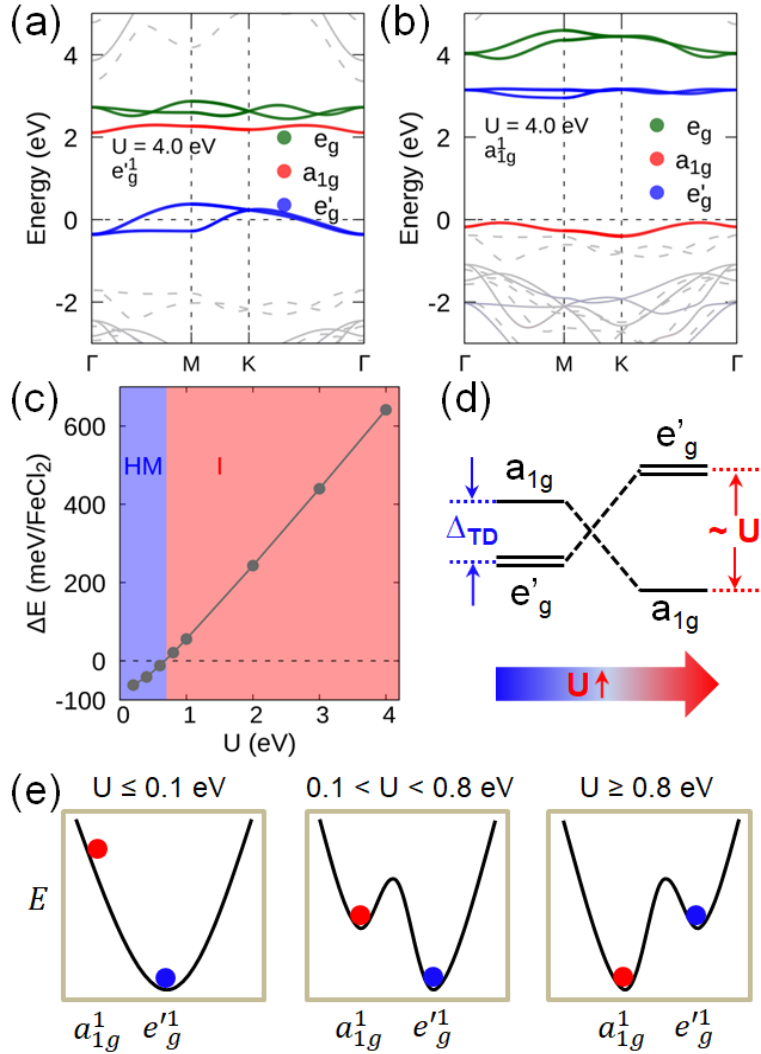

**Figure 2.** (a) Band structure calculated by spin-unrestricted GGA + $U$ (4 eV) method. The initial electron population is $e_g'^1 a_{1g}^0 e_g^0$, leading to the final $e_g'^1$ state. (b) Same as (a) but starts from initial electron population $e_g'^0 a_{1g}^1 e_g^0$, leading to the final $a_{1g}^1$ state. (c) Energy difference between $e_g'^1$ and $a_{1g}^1$ states as a function of $U$, $\Delta E = E(e_g'^1) - E(a_{1g}^1)$. HM stands for half-metal. I stands for insulator. (d) Schematic plot shows the competing effect between trigonal distortion $\Delta_{TD}$ and on-site Hubbard $U$. Orbital ordering can be reversed by sufficient $U$. (e) Schematic plot showing the potential surfaces at different $U$.

Based on the above analysis, we see that crystal field (including Jahn-Teller effect, *i.e.*, local trigonal distortion here) and electron correlation act as competing effects in determining the ground state of monolayer FeCl$_2$. Orbital ordering of $a_{1g}$-$e_g'$ can be switched by electron correlation. As schematized in Figure 2(d), without $U$ (or small $U$), local trigonal distortion splits $t_{2g}$ triplet into higher-lying $a_{1g}$ singlet and lower-lying $e_g'$ doublet. The $a_{1g}$-$e_g'$ gap is determined by the trigonal distortion $\Delta_{TD}$, which is typically tens to hundreds of meV. Further considering the electron correlation featured by the on-site Hubbard $U$, Coulomb repulsion raises up the energy of partially filled $e_g'$ orbitals. Sufficient $U$ (here ~ 0.8 eV) will reverse the orbital ordering, resulting in metal-insulator transition. Typical $U$ in 3$d$ transition compounds is about several eV (comparable to the magnitude of octahedral crystal field), much larger than $\Delta_{TD}$. Thus, the final gap is almost determined by the on-site Hubbard $U$. From 3$d$ to 4$d$ and 5$d$ transition compounds, although the electron correlation is reduced, the typical on-site Hubbard $U$ is still of the magnitude of eV. We expect the aforementioned conclusion still holds.

**Mechanism II: Degeneracy lifted by SOC.** Spin-orbit coupling is another important ingredient in the determination of electronic properties, which may induce a plethora of nontrivial effects such as topological phase transition. We next perform GGA + $U$ + SOC calculations to study the SOC effect. Same as the previous procedures, we start from two configurations of occupation, namely, $e_g'^1$ and $a_{1g}^1$. Owing to the unquenched orbital momentum, $e_g'$ orbitals are SOC-active states, which can form complex orbitals with $l_z = \pm 1$, breaking the doubly degeneracy of $e_g'$. In this case, SOC entails the half-metallic $e_g'^1$ state into an insulating state [14]. In the spin-down channel, SOC lowers $l_z = 1$ (labeled as $e'_{g+}$) state, and the $l_z = -1$ (labeled as $e'_{g-}$) state is pushed away from the Fermi level, leading to the final $e'^1_{g+}e'^0_{g-}a^0_{1g}e^0_g$ population (labeled as $e'^1_{g+}$). Figure 3(a) shows this mechanism exactly (compare with Figure 2(a)). It is worthwhile noticing that (at least most of) the large splitting between $e'_{g+}$ and $e'_{g-}$ is not solely contributed by SOC but also by electron correlation ($U = 4$ eV). Actually, at small $U$ (0.1 – 0.4 eV), the $e'_{g+}$-$e'_{g-}$ splitting is negligible. Band structures calculated with these $U$ still exhibit half-metallic characteristic (see Figure S5(a) to S5(c)). However, from $U = 0.8$ to 4 eV, cooperative effects of SOC and electron correlation gaps $e'_{g+}$ and $e'_{g-}$ states, making monolayer FeCl$_2$ a spin-orbit assisted Mott insulator, just like the famous iridate Sr$_2$IrO$_4$ [11]. Although in light elements, such as Fe and Cl, intrinsic SOC is much lower than that in 5$d$ transition metal iridium, the much stronger electron correlation becomes substantial for gap opening. Subtle balance between SOC and Hubbard correlation entails monolayer FeCl$_2$ preferring insulating other than half-metallic state.

At variance with $e'^1_g$, Figure 3(b) shows the band structure calculated staring from $a^1_{1g}$ (also see Figure S6). As aforementioned, $a^1_{1g}$ is a spin-only state, further considering the SOC effect doesn't bring about significant band modification. Tiny splitting can be observed in the first and second conduction bands ($e'_{g+}$-$e'_{g-}$ splitting). Since both $e'_{g+}$ and $e'_{g-}$ levels are unoccupied, the splitting between the two levels is dictated by SOC effects. Again, to ascertain the ground state of monolayer FeCl$_2$, it is necessary to compare the energy of $e'^1_{g+}$ and $a^1_{1g}$. As illustrated in Figure 3(c), monolayer FeCl$_2$ prefers $e'^1_{g+}$ with $U$ smaller than 1 eV, while for larger $U$, it favors $a^1_{1g}$. The half-metal-insulator transition occurs at $U \sim 0.7$ eV. Energy difference between $e'^1_{g+}$ and $a^1_{1g}$ converges progressively with the increase of $U$, which can be ascribed to the similar insulating band structures, regardless of different orbital ordering, in sharp contrast to $e'^1_g$ and $a^1_{1g}$ (Figure 2(c)).

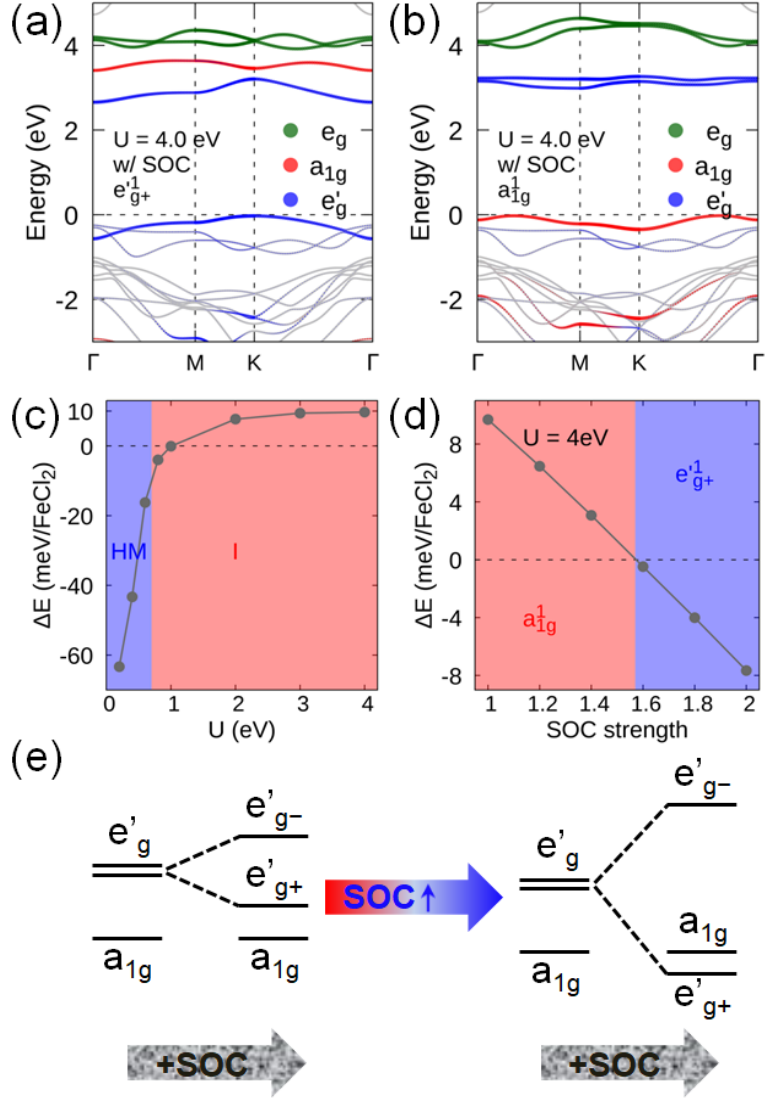

**Figure 3.** (a) Band structure calculated by GGA + $U$ (4 eV) + SOC method. The initial electron population is $e_g'^1$, leading to the final $e_{g+}'^1$ state. (b) Same as (a) but starts from initial electron population $a_{1g}^1$, leading to the final $a_{1g}^1$ state. (c) Energy difference between $e_{g+}'^1$ and $a_{1g}^1$ states as a function of $U$, $\Delta E = E(e_{g+}'^1) - E(a_{1g}^1)$. HM stands for half-metal. I stands for insulator. (d) Energy difference between $e_{g+}'^1$ and $a_{1g}^1$ states as a function of SOC strength. (e) Schematic plot shows the orbital ordering can be reversed by sufficient magnitude of SOC.

Meanwhile, the strength SOC can serve as another knob to tune the orbital ordering of monolayer FeCl$_2$. Figure 3(d) shows that at $U = 4$ eV, $a_{1g}^1$ can transit to $e_{g+}'^1$ when the magnitude of SOC is as strong as 1.6 times of the original strength. In particular, SOC lowers $e_{g+}'$ state of FeCl$_2$ but hardly affects $a_{1g}$ state, gradually enhanced SOC strength modifies $a_{1g}$-$e_{g+}'$ ordering, as schematized in Figure 3(e). Due to weaker electron correlation and stronger SOC, this phenomenon is ready to occur in 5$d$ transition metal compounds, such as in Sr$_3$NiIrO$_6$, orbital ordering induced by crystal-field eventually reversed by strong SOC of iridium [32].

**Mechanism III: Further symmetry-breaking energy-lowering distortions.** At last, we discuss the lift of degeneracy by the further structural distortions. Notice that in $e_g'^1$ without SOC, spin-down $e_g'$ doublet encompasses the Fermi level, which tends to result in symmetry-breaking Jahn-Teller distortions to lower the total energy. Such effect has been widely studied in transition-metal perovskite oxides, dubbed as one important gap-opening mechanism [33]. Possible symmetry-breaking distortions include octahedrons tilting, rotation, breathing and so on. So we carry out DFT+$U$ (4 eV) calculations starting from $e_g'^1$ and relax the structure without any enforced symmetry constrains (lattice constant is fixed to experimental value 3.603 Å, see Method for details). As anticipated, the relaxed structure lowers its symmetry with the final space group C2/m. Meanwhile, symmetry-lowering lifts the degeneracy of $e_g'$ and $e_g$ orbitals, leading to metal-insulator transition (Figure 4(a)). Such state is 675 meV lower than $e_g'^1$ and 33 meV lower than $a_{1g}^1$, respectively. Furthermore, we construct a 4 × 4 × 1 supercell to relax the structure with more degree of freedom. Without any symmetry contrains, the relaxed structure possesses space group P1. Figure 4(b) shows the effective band structure (EBS) of FeCl$_2$ supercell. Apart form the band broadening induced by structural disorder, we can see five singlely-degenerated $d$ orbital states between −0.5 to 5 eV obviously. Thus, apart from the orbital ordering mechanism, half-metallicity of FeCl$_2$ may also be excluded by possible structural distortions. In other words, "symmetry guaranteed" half-metal is somehow fragile when the half-metallicity is created by the partial filling of symmetry enforced degeneracy states. Various symmetry-breaking energy-lowering distortions tend to lift the degeneracy, resulting in metal-insulator transition.

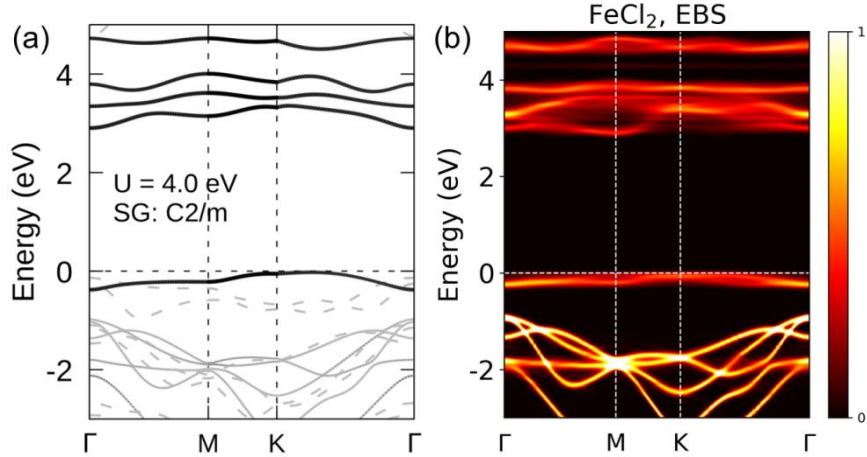

Figure 4. (a) Band structure of FeCl$_2$ with space group C2/m calculated by GGA + $U$ (4 eV) method. The spin-down Fe 3$d$ orbitals are highlighted by black. Spin-up and spin-down channels are indicated by dashed and solid lines, respectively. (b) Effective band structure of FeCl$_2$ 4×4×1 supercell with space group P1. The color scale represents the spectral weight of the $\bm{k}$ character for the primitive Brillouin zone. For clarity, we only show spin-down channel here.

## IV. CONCLUSION

In summary, exemplified by monolayer FeCl$_2$, our calculations demonstrate three mechanisms leading "half-metallic" FeCl$_2$ to be an insulator. Firstly, electron correlation reverses crystal field levels, rendering FeCl$_2$ a Mott insulator with the spin-down electron occupying $a_{1g}$ singlet. Bandgap between $a_{1g}$ and $e_g'$ can be ascribed to magnitude of Hubbard correlation. Secondly, SOC splits unoccupied $e_g'$ doublet without leading considerable modification of the band structure, making FeCl$_2$ as a SOC-assisted Mott insulator. Correlation and SOC act cooperatively to suppress the half-metallicity of monolayer FeCl$_2$. Gradually enhanced SOC will reverse $a_{1g}$-$e_{g+}'$ ordering but maintains the insulation, which may occur in 4$d$ and 5$d$ 2D materials. Thirdly, further symmetry-breaking structural distortions tend to lift the double degeneracy of $e_g'$, leading the metal-insulator transition of FeCl$_2$. Our results indicate that similar to transition metal compounds, charge, spin and orbital degrees of freedom can induce fruitful exotic effects in 2D magnets, and should be treated properly. Nowadays, high throughput simulation accelerates the process of material design. However, for complex quantum materials, which are hard to handle by naïve DFT calculations, extra filters should be added to avoid plausible predictions of fantasy materials.

**Acknowledgements**

This work was supported by National Key R&D Program of China (Grant No. 2020YFA0308900), the National Natural Science Foundation of China (Grant No. 12004163), Postdoctoral Fellowship of Shenzhen (K21207505), Guangdong Provincial Key Laboratory for Computational Science and Material Design (Grant No. 2019B030301001), the Shenzhen Science and Technology Program (Grant No. KQTD20190929173815000) and Center for Computational Science and Engineering of Southern University of Science and Technology.

**References**

[1]  B. Huang, G. Clark, E. Navarro-Moratalla, D. R. Klein, R. Cheng, K. L. Seyler, D. Zhong, E. Schmidgall, M. A. McGuire, D. H. Cobden, W. Yao, D. Xiao, P. Jarillo-Herrero, and X. Xu, Nature **546**, 270 (2017).

[2]  C. Gong, L. Li, Z. Li, H. Ji, A. Stern, Y. Xia, T. Cao, W. Bao, C. Wang, Y. Wang, Z. Q. Qiu, R. J. Cava, S. G. Louie, J. Xia, and X. Zhang, Nature **546**, 265 (2017).

[3]  Y. Deng, Y. Yu, Y. Song, J. Zhang, N. Z. Wang, Z. Sun, Y. Yi, Y. Z. Wu, S. Wu, J. Zhu, J. Wang, X. H. Chen, and Y. Zhang, Nature **563**, 94 (2018).

[4]  Z.-X. Shen, X. Bo, K. Cao, X. Wan, and L. He, Phys. Rev. B **103**, 085102 (2021).

[5]  Y. Tokura and N. Nagaosa, Science **288**, 462 (2000).

[6]  E. Dagotto, T. Hotta, and A. Moreo, Phys. Rep. **344**, 1 (2001).

[7]  E. Dagotto, Science **309**, 257 (2005).

[8]  C. Huang, F. Wu, S. Yu, P. Jena, and E. Kan, Phys. Chem. Chem. Phys. **22**, 512 (2020).

[9]  Y.-P. Wang and M.-Q. Long, Phys. Rev. B **101**, 024411 (2020).

[10] K. Yang, F. Fan, H. Wang, D. I. Khomskii, and H. Wu, Phys. Rev. B **101**, 100402 (2020).

[11] B. J. Kim, H. Jin, S. J. Moon, J. Y. Kim, B. G. Park, C. S. Leem, J. Yu, T. W. Noh, C. Kim, S. J. Oh, J. H. Park, V. Durairaj, G. Cao, and E. Rotenberg, Phys. Rev. Lett. **101**, 076402 (2008).

[12] X. Zhou, B. Brzostowski, A. Durajski, M. Liu, J. Xiang, T. Jiang, Z. Wang, S. Chen, P. Li, Z. Zhong, A. Drzewiński, M. Jarosik, R. Szczęśniak, T. Lai, D. Guo, and D. Zhong, J. Phys. Chem. C **124**, 9416 (2020).

[13] S. Cai, F. Yang, and C. Gao, Nanoscale **12**, 16041 (2020).

[14] A. S. Botana and M. R. Norman, Phys. Rev. Mater. **3**, 044001 (2019).

[15] E. Torun, H. Sahin, S. K. Singh, and F. M. Peeters, Appl. Phys. Lett. **106**, 192404 (2015).

[16] M. Ashton, D. Gluhovic, S. B. Sinnott, J. Guo, D. A. Stewart, and R. G. Hennig, Nano Lett. **17**, 5251 (2017).

[17] Y. Feng, X. Wu, J. Han, and G. Gao, J. Mater. Chem. C **6**, 4087 (2018).

[18] E. Ceyhan, M. Yagmurcukardes, F. M. Peeters, and H. Sahin, Phys. Rev. B **103**, 014106 (2021).

[19] R. K. Ghosh, A. Jose, and G. Kumari, Phys. Rev. B **103**, 054409 (2021).

[20] G. Kresse and J. Furthmüller, Phys. Rev. B **54**, 11169 (1996).

[21] P. Hohenberg and W. Kohn, Phys. Rev. **136**, B864 (1964).

[22] W. Kohn and L. J. Sham, Phys. Rev. **140**, A1133 (1965).

[23] J. P. Perdew, K. Burke, and M. Ernzerhof, Phys. Rev. Lett. **77**, 3865 (1996).

[24] G. Kresse and D. Joubert, Phys. Rev. B **59**, 1758 (1999).

[25] H. J. Monkhorst and J. D. Pack, Phys. Rev. B **13**, 5188 (1976).

[26] S. L. Dudarev, G. A. Botton, S. Y. Savrasov, C. J. Humphreys, and A. P. Sutton, Phys. Rev. B **57**, 1505 (1998).

[27] M. K. Wilkinson, J. W. Cable, E. O. Wollan, and W. C. Koehler, Phys. Rev. **113**, 497 (1959).

[28] P. V. C. Medeiros, S. Stafström, and J. Björk, Phys. Rev. B **89**, 041407 (2014).

[29] P. V. C. Medeiros, S. S. Tsirkin, S. Stafström, and J. Björk, Phys. Rev. B **91**, 041116 (2015).

[30] J. P. Allen and G. W. Watson, Phys. Chem. Chem. Phys. **16**, 21016 (2014).

[31] B. Dorado, B. Amadon, M. Freyss, and M. Bertolus, Phys. Rev. B **79**, 235125 (2009).

[32] X. Ou and H. Wu, Sci. Rep. **4**, 4609 (2014).

[33] J. Varignon, M. Bibes, and A. Zunger, Nat. Commun. **10**, 1658 (2019).

[34] K. Momma and F. Izumi, J. Appl. Cryst. **44**, 1272 (2011).

[35] V. Wang, N. Xu, J. Liu, G. Tang, and W. Geng, arXiv: 1908.08269 (2019).

[36] L. Hu, C. Xie, S. J. Zhu, M. Zhu, R. H. Wei, X. W. Tang, W. J. Lu, W. H. Song, J. M. Dai, R. R. Zhang, C. J. Zhang, X. B. Zhu, and Y. P. Sun, Phys. Rev. B **103**, 085119 (2021).

[37] S. Landron and M.-B. Lepetit, Phys. Rev. B **77**, 125106 (2008).

[38] H. Wu, C. F. Chang, O. Schumann, Z. Hu, J. C. Cezar, T. Burnus, N. Hollmann, N. B. Brookes, A. Tanaka, M. Braden, L. H. Tjeng, and D. I. Khomskii, Phys. Rev. B **84**, 155126 (2011).

[39] F. Aryasetiawan, K. Karlsson, O. Jepsen, and U. Schönberger, Phys. Rev. B **74**, 125106 (2006).